\documentclass[conference]{IEEEtran}
\IEEEoverridecommandlockouts
\usepackage{cite}
\usepackage{amsmath,amssymb,amsfonts}
\usepackage{algorithmic}
\usepackage{graphicx}
\usepackage{textcomp}
\usepackage{xcolor}

\usepackage{xurl}
\usepackage{booktabs}
\usepackage{color}
\usepackage{url}
\usepackage{multirow}
\usepackage{diagbox}
\usepackage{float}
\usepackage{subfig}
\usepackage{enumitem}
\def\BibTeX{{\rm B\kern-.05em{\sc i\kern-.025em b}\kern-.08em
    T\kern-.1667em\lower.7ex\hbox{E}\kern-.125emX}}
\begin{document}

\title{Generative Machine Listener\\
}


\author{\IEEEauthorblockN{Guanxin Jiang}
	\IEEEauthorblockA{\textit{Sound Tech Research} \\
		\textit{Dolby Germany GmbH}\\
		 guanxin.jiang@dolby.com
		}
	\and
	\IEEEauthorblockN{Lars Villemoes}
	\IEEEauthorblockA{\textit{Sound Tech Research} \\
		\textit{Dolby Sweden AB}\\
		 lars.villemoes@dolby.com
		}
	\and
	\IEEEauthorblockN{Arijit Biswas}
	\IEEEauthorblockA{\textit{Sound Tech Research} \\
		\textit{Dolby Germany GmbH}\\
		 arijit.biswas@dolby.com
		}
}

\maketitle

\begin{abstract}
We show how a neural network can be trained on individual intrusive
listening test scores to predict a distribution of scores for each pair of
reference and coded input stereo or binaural signals. We nickname this method the Generative Machine Listener (GML), as it is capable of generating an arbitrary amount of simulated listening test data. Compared to a baseline system
using regression over mean scores, we observe lower outlier ratios (OR)
for the mean score predictions, and obtain easy access to the prediction of 
confidence intervals (CI). The introduction of data augmentation techniques 
from the image domain results in a significant increase in CI prediction accuracy 
as well as Pearson and Spearman rank correlation of mean scores.
\end{abstract}

\begin{IEEEkeywords}
Objective audio quality metrics, audio coding, deep learning, generative modeling 
\end{IEEEkeywords}

\section{Introduction}
\label{section:Introduction}

It has been found challenging by the authors to train a neural network to directly output predictions 
of both mean values and confidence intervals (CI) of listener scores. CIs are one way to represent how good the estimated subjective scores are in a listening test. A narrower CI indicates a more precise estimate, while a wider CI indicates a less precise estimate. Subjective CIs are used to judge the statistical significance of the differences between systems under test, e.g., different codecs. From the user satisfaction point of view, a system with a high mean quality score, but with a small CI is much preferable over a large CI. Thus, if an audio quality metric is also able to predict CI, it adds value to automatic quality evaluation. 

The modeling of data uncertainty has been pursued with histogram matching objectives on both non-intrusive 
and intrusive quality scores for images~\cite{NIMA} 
and non-intrusive Mean Opinion Score (MOS) for audio~\cite{faridee22_MOSdistr}. However, no study of CI was reported for the latter. (Note that estimation of the uncertainty of the model itself is a different topic, as recently quantified via a bootstrapping approach in the video domain~\cite{VMAF_CI}.)  

In this paper, we apply the methods of generative modeling to obtain straightforward predictions of CI. The task of the Generative Machine Listener (GML) is to efficiently simulate an arbitrary number of listener scores for a given input signal pair. 
As opposed to regression on mean scores, the training based on the maximum likelihood principle is influenced in proportion to the human effort even for listening test data with a varying number of listeners.

We also exploit data augmentation, in particular, a variant of MixUp~\cite{zhang2018mixup}, called the CutMix~\cite{CutMix}. These are popular data augmentation techniques from the image domain and are used heavily in classification tasks. With MixUp, authors reported robustness when learning from corrupt labels or adversarial examples. CutMix further improved upon the robustness of the unseen MixUp samples and alleviated the overconfidence of the model. These techniques were also adapted for speech and audio, but mostly for classification~\cite{kim21c_SpecMix} and recognition~\cite{park19e_SpecAugment} tasks, but not for audio quality prediction or generative modeling. 

The paper is organized as follows. Section~\ref{section:GML} describes the proposed generative modeling concept for predicting audio quality.  In Section~\ref{section:Datasets}, the data used for model training and evaluation are introduced. The related experimental results and analysis are given in Section~\ref{section:Results}, and finally, the conclusion is drawn in Section~\ref{section:ConclusionDiscussion}.

\section{Generative Machine Listener}
\label{section:GML}

Given a reference signal $x$ and a signal under test $y$, the GML model provides a 
probability distribution of scores $s$ for $y$ by a parametrized probability density function,
\begin{equation}
	p_\theta(s \vert x,y).
\end{equation}
The generative aspect of the model is that a listening test with $N$ listeners can in principle be simulated by 
sampling the model $N$ times. However, by using explicit output distributions, the desired statistics can also be derived directly from the parameters. 

\begin{figure}[t]
	\centering
	\includegraphics[width=1.0\linewidth]{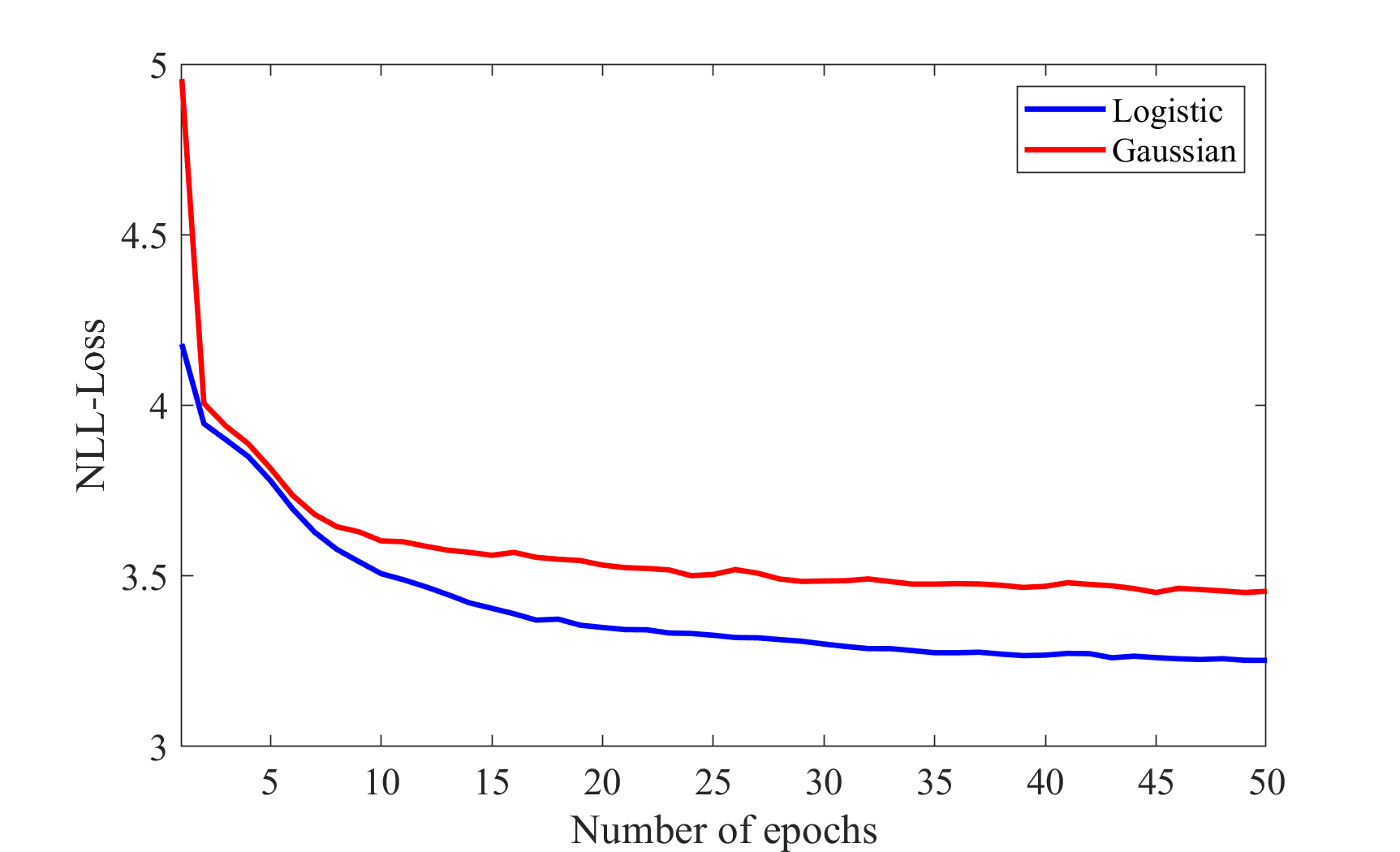} 
	\caption{NLL training losses with Logistic (blue) and Gaussian distribution (red).}
	\label{fig:GaussianVsLogistic}
\end{figure}

We consider two example distributions, gaussian and logistic.
Given $x$ and $y$, the model outputs a mean value $\mu$ and the logarithm of the gaussian standard deviation $\sigma$ or the logistic scale parameter $a$. For training of the model parameters $\theta$, 
we use the negative log-likelihood (NLL) loss, whose
contribution for each triplet $(x,y,s)$ is given by either \eqref{lossgauss} or \eqref{losslogistic}:
\begin{align}
	\label{lossgauss}
	L_\text{gaussian} &= \log{ (\sqrt{2\pi} \sigma)} + \frac{(s-\mu)^2}{2\sigma^2},
	\\
	\label{losslogistic}
	L_\text{logistic} &= \log (4a) + 2 \log \operatorname{sech} \bigl(  \frac{s-\mu}{2a} \bigr).
\end{align}
The gaussian distribution may seem like the most natural choice here, but we include the logistic alternative due to its previous usage as scalar distribution for predictive generative modeling \cite{salimans2017pixelcnn}.
As a side note, there is a qualitative relation to the smooth L1-loss~\cite{SmoothL1LossPytorch} frequently used for regression,
\begin{equation}
	\label{losssmoothl1}
	L_\text{smooth} = 
	\begin{cases} 
		\tfrac12 (s-\mu)^2,  & |s-\mu| < 1 ; \\
		|s-\mu| - \tfrac12, & \text{otherwise.}
	\end{cases}
\end{equation}
Specifically, for fixed scales in the range $a \in [1/\sqrt{2} ,1]$, one observes similarities between $L_\text{logistic}$ and $L_\text{smooth}$ up to constant shifts. However, our aim here is density modeling with two-parameter models and we have found that the logistic distribution performs best in terms of NLL loss for the task at hand (Figure~\ref{fig:GaussianVsLogistic}). 

The statistics we care about are mean and CI. For the mean, we use the model output $\mu$ directly. For consistency with the test sets, we compute t-distribution-based 95\% confidence intervals given by the number of listeners $N$ in each test and the standard deviation provided by the model. The value of this standard deviation is $\pi a/\sqrt{3}$ for the logistic distribution.


\section{Datasets}
\label{section:Datasets}

\subsection{Training set}
\label{subsection:trainingset}

We used our internal corpus of Multiple Stimuli with Hidden Reference and Anchor (MUSHRA)~\cite{MUSHRA} listening tests at a 48 kHz sample rate, all auditioned with headphones. Each of these tests included an unencoded hidden reference, the 3.5~kHz and 7~kHz low-pass filtered versions of the unencoded signals, and one or more coded signals. The stereo codecs used in the training set were AAC~\cite{AAC}, HE-AAC v1 and v2~\cite{HE-AAC}, and Dolby AC-4~\cite{ac4_IEEE} spanning a wide range of bitrates. We included all content types but excluded dedicated speech codecs. Similarly, binaural listening tests included object-based immersive audio content (e.g., Dolby Atmos~\cite{Robinson_Atmos}) coded with AC-4 Immersive Stereo (IMS)~\cite{IMS_whitepaper} and with DD+JOC~\cite{purnhagen2016immersive} and AC-4 A-JOC~\cite{purnhagen2016immersive},\cite{ac4_IEEE} rendered to binaural after decoding. We used binaural renditions of Dolby Atmos as the uncoded reference. We also included tests with 3GPP IVAS codec~\cite{Bruhn_IVAS} for coding (first and higher order) ambisonic signals, where the references and decoded ambisonic signals are rendered to binaural~\cite{IVASgithub}. In total, we used 67,505 subjective scores.

Note that the listening test excerpts can be of different lengths. Therefore, we extended the smaller excerpts to the maximum length by zero-padding them on both sides. Furthermore, as proven to be beneficial~\cite{StereoInSE-NET}, we expanded the training dataset by swapping the left- and right-channel of all the audio signals, but preserved the assigned quality labels.

\subsection{Test sets}
\label{subsection:testset}

We benchmarked the prediction accuracy of GML against subjective listening scores from the Unified Speech and Audio Coding (USAC)~\cite{MPEG-USAC} verification listening tests~\cite{usac_lt},\cite{USAC}. These comprehensive tests contain 24 excerpts coded with USAC, HE-AAC, and AMR-WB+ with bitrates ranging from 8~kb/s mono to 96~kb/s stereo. It consists of three separate listening tests: mono at low bitrates and stereo at both low and high bitrates. The number of listeners (after screening) in the mono, stereo low bitrate, and stereo high bitrate tests were 66, 44, and 28, respectively. All tests were MUSHRA tests, with a 0 to 100 quality scale, where a higher score implies better quality. Note that we included the mono listening test because we also wished to evaluate the accuracy of the GML when trained with only coded stereo and binaural listening tests.

Since we lacked access to associated subjective scores from a relevant binaural listening test~\cite{rudzki2019perceptual}, we used two internal MUSHRA listening tests: Binaural Test-1 and 2, auditioned over headphones by 9 and 11 subjects, respectively. Test-1 consists of 11 excerpts coded with two variants of DD+JOC at 448 kb/s and two variants of IMS at 256 kb/s. Test-2 consists of 12 excerpts coded with IMS at 64, 112, and two variants of 256 kb/s. None of the excerpts were used in the training.

In all test sets, the computation of 95\% confidence intervals was based on the t-distribution. 


\section{Experiments and Results}
\label{section:Results}

\subsection{Model architecture}
\label{subsection:model_arch}

The concept of GML is generic and the principle can be applied to any Deep Neural Network (DNN)-based audio quality prediction model to make it ``generative''. In the proposed GML, we utilized the DNN-based stereo model~\cite{StereoInSE-NET} as the backbone. For details about the architecture of the DNN-based stereo model, interested readers are referred to Figure 3 and Table 1 in~\cite{StereoInSE-NET}, and the motivation behind such a convolutional architecture design are described in~\cite{InSE-NET}. Thus, the input to GML is the Gammatone spectrograms of reference-coded (ref.-cod.) pairs for left ($L$) and right ($R$), mid ($M = (L+R)/2$), and side ($S = (L-R)/2$) channels. The output stage of~\cite{StereoInSE-NET} is also augmented (i.e., modified))to two dimensions for providing a distribution of MUSHRA scores as described in Section~\ref{section:GML}. Furthermore, unlike~\cite{StereoInSE-NET}, given a ref.-cod. pairs of audios, we utilize individual listener scores for training (as opposed to mean subjective score as the training target). Our proposed GML has 15.25M parameters, only a meager $0.0033\%$ more than its non-generative counterpart~\cite{StereoInSE-NET}.

\subsection{Training configuration}
\label{subsection:training_config}

The training dataset was first normalized and partitioned randomly into $80\%$ for training and $20\%$ for validation. A 5-fold cross-validation is applied to ensure that the model could make full use of the listening scores. The setup is implemented with PyTorch and was trained for 10 epochs for each fold (i.e., 50 epochs in total) on an Nvidia A100 GPU with Adam optimizer. We kept the optimal kernel sizes (as described in~\cite{StereoInSE-NET}), a learning rate of $10^{-4}$, batch size of 8, and trained the model from scratch with the default PyTorch initializer~\cite{He_2015}.  The model is trained with NLL loss and evaluated with the following criteria: NLL loss, Pearson's correlation coefficient ($R_p$), and Spearman's correlation coefficient ($R_s$). The correlation coefficients $R_p$ and $R_s$ are used to measure the linear and monotonic relationships between two continuous variables, respectively. Note that $R_s$ is based on the ranked values for each variable rather than the raw data itself, and it is used to measure the rank preservation. For exact definitions of $R_p$ and $R_s$, readers are referred to~\cite{RpRscorr}.

\subsection{Data augmentation}
\label{subsection:dataaugmentation}

In addition to swapping of left and right channels (Section~\ref{subsection:trainingset}), we also explored the vanilla MixUp~\cite{zhang2018mixup} and its successor, the CutMix~\cite{CutMix} as data augmentation strategies. The general idea of MixUp translated to the context of our work would be to blend Gammatone spectrograms of two different signals, and their associated quality labels by varying a random $\lambda$ drawn from a beta distribution $\mathbf{B}(\alpha,\beta)$. After trying out MixUp and observing that the training and validation losses decayed more smoothly, we explored vanilla CutMix and report the results here. CutMix randomly cuts out and attaches a part of a spectrogram to another spectrogram. It applies a randomly generated mask for cutting out a spectrogram region, pastes it randomly to another spectrogram region, and creates a new (CutMixed) spectrogram with its associated quality. The ratio of the cut-out area-to-remaining area of the spectrogram is determined by the hyperparameter $\lambda \sim \mathbf{B}(\alpha,\beta)$. The associated quality score ($\tilde{y}$) of the CutMixed spectrogram is obtained by a weighted linear combination of per listener quality scores ($y_{A}$ and $y_{B}$) of two involved spectrograms as
\begin{equation}
	\tilde{y} = \lambda y_{A} + (1-\lambda) y_{B}. 
\end{equation}
The operation is done on the fly (per batch) on the GPU. Typically, $\alpha = \beta$, is a hyperparameter that one needs to tune on the validation set. We chose $\alpha = 0.7$. For algorithmic details of the CutMix, interested readers are referred to~\cite{CutMix}. We followed the same algorithm, but repurposed, trained, and tuned it for our application. 

\subsection{GML Benchmarking}
\label{subsection:benchmark}

\begin{table*}[t]
	\setlength\tabcolsep{5pt}
	\centering
	\caption{Performance of GML on USAC verification and binaural listening tests. The table shows (a) correlation coefficients ($R_p$ and $R_s$)$\uparrow$ and outlier ratios (OR)$\downarrow$ between predicted and subjective mean MUSHRA scores, and (b) correlation coefficients and root mean squared error (RMSE)$\downarrow$ between predicted and subjective CI.}
	
	\vspace{-0.4cm}
	\begin{center}
		\scriptsize{
			\begin{tabular}{|l||c|c|r|c|c|r|c|c|r|c|c|r|c|c|r|}
				\hline
				\multirow{2}{*}{\backslashbox{\textbf{Model}}{\textbf{Metric}}}   & \multicolumn{3}{c|}{\textbf{Mono Bitrates}} & \multicolumn{3}{c|}{\textbf{Stereo Low Bitrates}}   & \multicolumn{3}{c|}{\textbf{Stereo High Bitrates}}      & \multicolumn{3}{c|}{\textbf{Binaural Test-1}}   & \multicolumn{3}{c|}{\textbf{Binaural Test-2}} \\ \cline{2-16} 
				& $\mathbf{R_p}$   & $\mathbf{R_s}$ & \multicolumn{1}{c|}{$\mathbf{OR}$} & $\mathbf{R_p}$   & $\mathbf{R_s}$ & \multicolumn{1}{c|}{$\mathbf{OR}$} & $\mathbf{R_p}$   & $\mathbf{R_s}$ & \multicolumn{1}{c|}{$\mathbf{OR}$} & $\mathbf{R_p}$   & $\mathbf{R_s}$ & \multicolumn{1}{c|}{$\mathbf{OR}$} & $\mathbf{R_p}$   & $\mathbf{R_s}$ & \multicolumn{1}{c|}{$\mathbf{OR}$}\\ \hline\hline
				\textbf{ViSQOL-v3}      & 0.81  & 0.84  & n.a.  & 0.77  & 0.78  & n.a.   & 0.82  & 0.90  & n.a.   & 0.90  & 0.93  & n.a.   & 0.96  & 0.85  & n.a. \\ \hline
				\textbf{Non-GML}        & 0.87  & 0.82  & 0.92 & 0.87  & 0.83  & 0.82  & \textbf{0.93}  & 0.93  & 0.78  & \textbf{0.98}  & \textbf{0.96}  & 0.27  & 0.98  & 0.89  & 0.77\\ \hline
				\textbf{GML}   & 0.84  & 0.80  & \textbf{0.75} & 0.82  & 0.75  & \textbf{0.63}  & 0.90  & 0.90  & 0.62  & 0.96  & 0.94  & 0.34  & \textbf{0.99}  & \textbf{0.95}  & \textbf{0.42}\\ \hline
				
				\textbf{GML w/ CutMix}  & \textbf{0.88}  & \textbf{0.88}  & 0.80 & \textbf{0.89}  & \textbf{0.86}  & 0.70  & 0.92  & \textbf{0.94}  & \textbf{0.56}  & \textbf{0.98}  & 0.95  & \textbf{0.19}  & 0.98  & 0.92  & 0.56\\ \hline
				\textbf{Non-GML /w CutMix}        & 0.87  & 0.83  & 0.87 & 0.87  & 0.80  & 0.80  & 0.90  & 0.89  & 0.78  & \textbf{0.98}  & 0.95  & 0.23  & \textbf{0.99}  & \textbf{0.95}  & 0.51\\ \hline
		\end{tabular}}
	\end{center}

	\subfloat[\label{tbl:mean}]
	\newline
	\vspace{-0.6cm}
	\begin{center}
		\scriptsize{
			\begin{tabular}{|l||c|c|r|c|c|r|c|c|r|c|c|r|c|c|r|}
				\hline
				\multirow{2}{*}{\backslashbox{\textbf{Model}}{\textbf{Metric}}}   & \multicolumn{3}{c|}{\textbf{Mono Bitrates}} & \multicolumn{3}{c|}{\textbf{Stereo Low Bitrates}}   & \multicolumn{3}{c|}{\textbf{Stereo High Bitrates}}      & \multicolumn{3}{c|}{\textbf{Binaural Test-1}}   & \multicolumn{3}{c|}{\textbf{Binaural Test-2}} \\ \cline{2-16} 
				& $\mathbf{R_p}$   & $\mathbf{R_s}$ & \multicolumn{1}{c|}{$\mathbf{RMSE}$} & $\mathbf{R_p}$   & $\mathbf{R_s}$ & \multicolumn{1}{c|}{$\mathbf{RMSE}$} & $\mathbf{R_p}$   & $\mathbf{R_s}$ & \multicolumn{1}{c|}{$\mathbf{RMSE}$} & $\mathbf{R_p}$   & $\mathbf{R_s}$ & \multicolumn{1}{c|}{$\mathbf{RMSE}$} & $\mathbf{R_p}$   & $\mathbf{R_s}$ & \multicolumn{1}{c|}{$\mathbf{RMSE}$}\\ \hline\hline
				\textbf{GML}            & 0.36  & 0.28 & 2.80 & 0.31  & 0.23  & 3.82 & 0.38  & 0.34  & 4.44 & 0.37  & 0.38  & 7.61 & 0.21  & 0.28  & 4.46 \\ \hline
				\textbf{GML w/ CutMix}  & \textbf{0.79}  & \textbf{0.44} & \textbf{0.87} & \textbf{0.80}  & \textbf{0.43}  & \textbf{1.13} & \textbf{0.78}  & \textbf{0.67}  & \textbf{1.50} & \textbf{0.70}  & \textbf{0.65}  & \textbf{3.20} & \textbf{0.76}  & \textbf{0.60}  & \textbf{2.25} \\ \hline
		\end{tabular}}
	\end{center}
	\subfloat[\label{tbl:CI}]
	\newline
\end{table*}

We benchmark GML against ViSQOL-v3 (operating in audio mode)~\cite{ViSQOLgithub}. Note that unlike GML, ViSQOL-v3 is not a DNN-based coded audio quality predictor. We use ViSQOL-v3 as a benchmark because it has been reported in \cite{Fraunhofer}, that out of all objective measures designed to evaluate codecs, ViSQOL shows the best correlation with subjective scores and achieves high and stable performance for all content types. In addition, we benchmark against our non-generative counterpart (non-GML), i.e., using the same base model as used in GML, but trained on the same dataset to predict the mean MUSHRA score with smooth L1-loss~\cite{SmoothL1LossPytorch}. In our studies we only considered the GML trained with the logistic distribution because the decay of training losses (Figure~\ref{fig:GaussianVsLogistic}) already indicates its superiority over the gaussian distribution. We used $R_s$ to measure the prediction monotonicity of the models and $R_p$ to measure the prediction linearity. For both $R_p$ and $R_s$, larger values denote better performance. Furthermore, we can also easily compute the outlier ratio (OR)~\cite{yi22b_outlierRatio} defined as the ratio of the number of outliers to the total number of excerpts. The predicted score is an outlier if it is greater than the 95\% CI of the subjective MUSHRA score. For OR, a lower value implies better performance. The accuracy of CI prediction is quantified in terms of $R_p$, $R_s$, and root mean squared error (RMSE), where a lower value implies better performance. The performance numbers for the test sets are listed in Table I, with best performing models indicated in bold. In Table I(a), we do not report OR with ViSQOL-v3 because the predicted MOS score is bounded between [1, 4.732]~\cite{InSE-NET},\cite{ViSQOLgithub}, and we are unaware of a suitable mapping to convert such a scale to MUSHRA before computing the OR. 

\begin{figure}[t]
	\centering
	\includegraphics[width=1.0\linewidth]{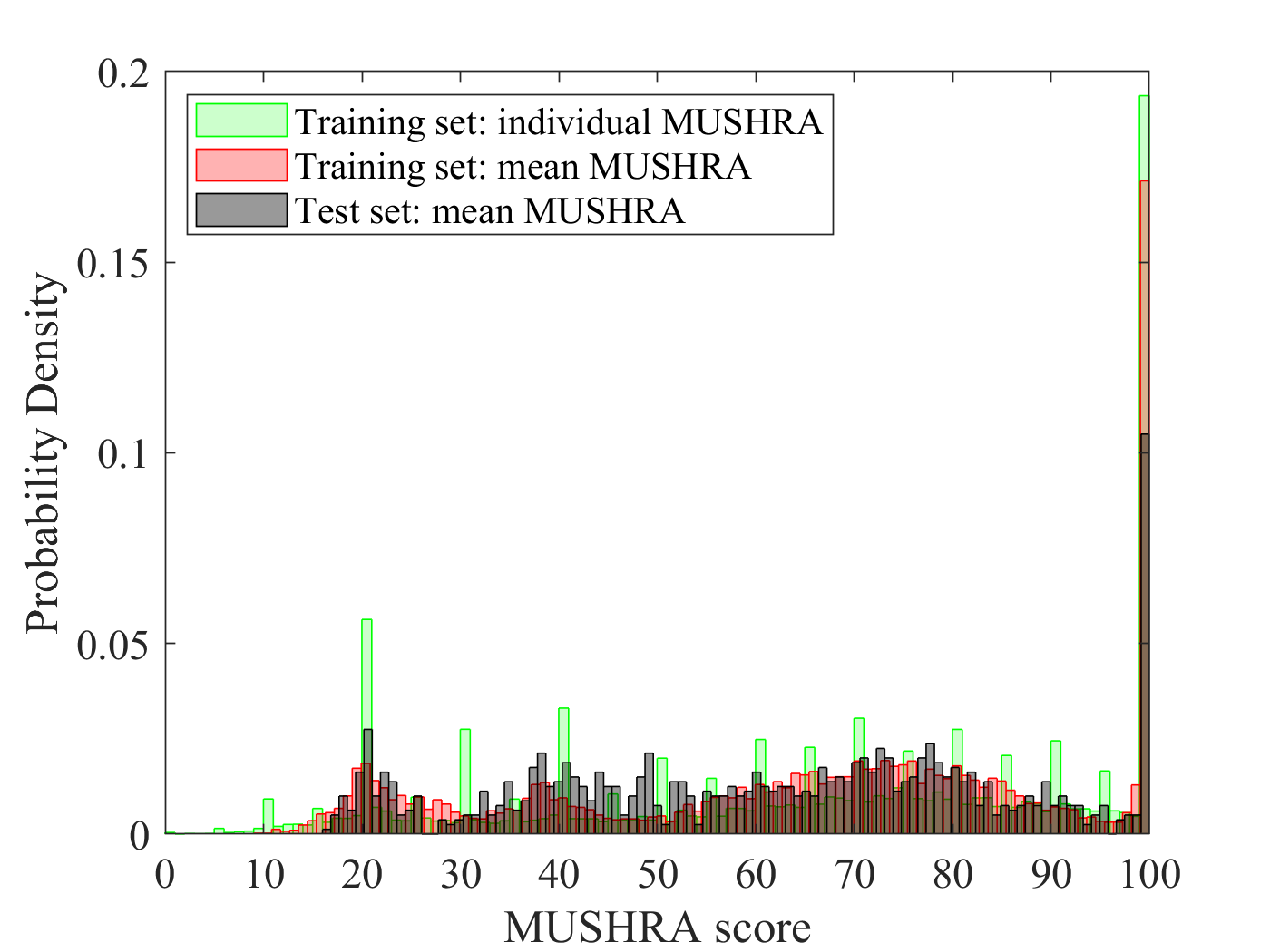} 
	\caption{Histogram of mean MUSHRA (training and test set) and individual MUSHRA scores (training set). Histograms of the training sets exclude data augmentation.}
	\label{fig:MushraDistribution}
\end{figure}

We observe across all test sets and evaluation metrics that GML trained with CutMix is the best-performing model followed by the GML trained without CutMix. Furthermore, when both the GMLs (i.e., trained without and with CutMix) are considered jointly, we can observe that OR is lower for all five tests hinting at robust prediction. For a couple of cases, the correlation numbers are worse than the non-GML. We argue this is due to the similarity in distribution between training and test sets (shown in Figure~\ref{fig:MushraDistribution}). Hence, to be fair to the GML, one needs to consider that we are training on individual subjective scores to predict a description of distribution (as opposed to training/testing for mean score), and its additional capability to predict the CI. With the GML trained with CutMix, we observe (Table I(b)) a significant improvement in $R_p$, $R_s$, and RMSE between predicted and subjective CI.

For the mono listening test, we evaluate a dual-mono (i.e., stereo with L = R) signal.  We can observe that even though none of our models were trained with mono listening tests, they display a strong $R_p$ compared to ViSQOL. Both $R_p$ and $R_s$ got significantly improved with GML trained with CutMix, suggesting improved robustness to unseen conditions with data augmentation.  

At this juncture, it is valid to also crosscheck any impact of training the non-GML model with CutMix, even though the non-GML model is not capable of predicting the CI. So, its prediction accuracy is reported in the last row in Table I(a). Note that for the non-GML, the quality score of the CutMixed spectrogram is obtained by a weighted linear combination of mean scores of two involved spectrograms, as opposed to per listener quality scores used in the GML. The optimal CutMix hyperparameter was also found to be $\alpha = 0.7$. After training with CutMix, the non-GML achieves the best correlation numbers for one listening test. However, as compared to GML, the added improvement of CutMix is much less. 

As can be seen from Table I(a), GML trained with Cutmix most often achieves the best (top-1) performance scores in terms of $R_p$ or $R_s$ or OR. For the cases where the model fails to achieve the best performance, it still achieves a top-2 performance score, except for $R_s$ and OR in Binaural Test-2, where it achieves a top-3 performance score. These results demonstrate the robustness of the GML model trained with CutMix regardless of test conditions.

\section{Conclusion and Discussion}
\label{section:ConclusionDiscussion}

We described GML, a novel generic concept for predicting both mean quality score and CI. The model is trained with individual subjective ratings, and we discovered that data-augmented training with CutMix significantly improves the accuracy. We speculate that the improvement in mean quality prediction accuracy is due to the smoothing (without any information loss) of the distribution of individual scores with linear interpolation, as well as due to a reduced impact of possible corrupt labels (e.g., due to listeners' skills and bias). Furthermore, deep learning-based models in general tend to be over-confident in making a prediction. Likewise, GML trained without CutMix also tends to be overconfident and predict a very narrow CI. However, when GML is trained with Cutmix, the model is forced to see more ``in-between'' examples, where the quality decision boundaries may be blurry. Such augmented data alleviated the overconfidence of the GML and significantly improved the CI prediction accuracy. 

Concerning mean score prediction only, we observe less benefit of CutMix for the non-GML model. Speculative reasons for this could be the following. First, for the task of mean quality prediction using the mean quality score as the training target, additional smoothing of mean scores with linear interpolation may not be beneficial. Second, identical CutMixed spectrograms will always have the same quality scores, whereas in the case of GML, since individual scores are combined, they may have different scores. Thus, non-GML trained with CutMix adds less diversity to the dataset.

The base DNN-based stereo model~\cite{StereoInSE-NET} in the GML was adapted from the image domain~\cite{InSE-NET}. Also, as mentioned, the data augmentation technique with CutMix was introduced in the image domain and is used heavily in image classification tasks. So, it is likely that the proposed generative concept presented here for audio quality modeling could be also applied to image and video quality assessment. 

Finally, the proposed GML enables faster prediction of both mean MUSHRA quality scores and CI. The proposed GML is a hybrid implementation in Python and PyTorch. The Gammatone spectrogram computation frontend is in Python and the DNN-based model is in PyTorch. Excluding the Gammatone spectrogram computation, the DNN model runs at 32.3x real-time on a CPU with the latest PyTorch version 2.0. Including the Gammatone spectrogram computation frontend, the GML runs at 1.2x real-time on a CPU, which is slightly faster than ViSQOL-v3 running at 1.1x real-time. However, note that ViSQOL-v3 consists of traditional signal processing-based blocks which are fully implemented in C++, but it was not designed to predict the CI. Furthermore, ViSQOL-v3 computes a pair of Gammatone spectrograms, whereas we compute four pairs of Gammatone spectrograms and feed them to the model. Therefore, even a slight advantage of the GML in computational efficiency complemented with improved audio quality prediction accuracy over ViSQOL-v3 and its ability to predict CI, is a positive outcome.

\bibliographystyle{IEEEtran}


\end{document}